\documentclass[aps,prl,twocolumn]{revtex4-1}
\usepackage{amsmath}
\usepackage[usenames]{color}
\usepackage{float}
\usepackage{graphicx}
\usepackage{hyperref}
\usepackage{wrapfig}
\usepackage{amssymb}
\usepackage{enumitem}
\usepackage{bm}
\usepackage{subfigure}
\usepackage{placeins}
\usepackage{gensymb}
\usepackage{xcolor}

\begin{document}

\title{A DMRG Study of Superconductivity in the Triangular Lattice Hubbard Model}

\author{Jordan Venderley and Eun-Ah Kim}
\affiliation{Cornell University, Ithaca, NY  14850}
\date{\today} 

\begin{abstract}
    With the discovery of strong coupling physics and superconductivity in Moir\'e superlattices, it's essential to have an understanding of strong coupling driven superconductivity in systems with trigonal symmetry. 
    The simplest lattice model with trigonal symmetry is the triangular lattice Hubbard model. Although the triangular lattice spin model is a heavily studied model in the context of frustration, studies of the hole-doped triangular lattice Hubbard model are rare. Here we use density matrix renormalization group (DMRG) to investigate the domininant superconducting channels in the hole-doped triangular lattice Hubbard model over a range of repulsive interaction strengths. 
    We find a clear transition from $p$-wave superconductivity at moderate on-site repulsion strength ($U/t = 2$) at filling above 1/4 ($n \sim 0.65$) to $d$-wave superconductivity at strong on-site repulsion strength ($U/t = 10$)  at filling below 1/4 ($n \sim 0.4$). 
    The unusual tunability that Moir\'e superlattices offer in controlling $U/t$ would open up the opportunity to realize this transition between $d$-wave and $p$-wave superconductivity.
    
\end{abstract}

\maketitle

\section{Introduction}

The discovery of superconductivity in van der Waals materials \cite{Pablo2018} with an (effective) triangular lattice structure has expanded the scope of correlation physics in triangular lattice models. Nevertheless, recent literature motivated by the experiments\cite{Pablo2018,pablo2,Young-Dean} have for the most part been limited to mean-field theories\cite{Xu-Balents,Dodaro-Kivelson,Fidrysiak-PhysRevB.98.085436} and perturbative renormalization group studies\cite{Isobe-Fu,Roy-2018arXiv180311190R}. Such approximations are unavoidable when one tries to capture the full band structure of Moir\'e superlattices such as magic angle twisted bilayer graphene, which require several bands and large unit-cells to accommodate the characteristic Moir\'e patterns that ultimately drive their correlated behavior. 

However, there is increasing awareness that the effective band structure of a wider class of Moir\'e superlattice systems is much simpler\cite{MacDonald2018,MacDonald-Wu2018}. In particular, Ref.~\cite{MacDonald2018}
showed a simple triangular lattice Hubbard model adequately captures hetero-transition metal dichalcogenide (TMD) bilayers. The observation of superconductivity and other strong coupling phenomena in these systems\cite{PhilipKim} motivates studying simple effective models with strong coupling approaches. 
In fact, \textcite{Scalettar-PhysRevB.97.235453} have already studied a three-band triangular lattice Hubbard model with modulated hopping using determinant quantum Monte Carlo in the context of Moir\'e superlattices. Similarly, \textcite{Fu2018}, studied a 2-orbital model on the honeycomb lattice with density matrix renormalization group (DMRG) in the insulating state.
In this work, we study strong coupling driven superconductivity and investigate how the strength of the repulsive interaction affects the pairing symmetry in the triangular lattice Hubbard model. 

Even before the discovery of superconductivity in twisted bilayer graphene, superconductivity in the organic salts \cite{Matsuda2008,Kanoda2008} and cobaltates\cite{Sasaki2003} had generated interest in strong interaction driven superconductivity in triangular lattice systems. Gutzwiller projected t-J models with smaller Hilbert space predicted $d+id$ pairing for light to moderate hole-dopings away from half-filling\cite{Ogata2004,Lee2004,Ran2016}. For the Hubbard model, \textcite{Moreno2013} used dynamical cluster quantum Monte Carlo to study strong repulsion with moderate hole-doping to find the dominant pairing susceptibility in the $d+id$ channel.
These 2D irreps are of theoretical interest because of the non-zero Chern number associated with their quasi-particle spectra and the corresponding ability to support Majorana bound states. In particular, the Majorana's are expected to be better protected in the (\textit{p}$\pm i$\textit{p}) paired state. 

 Here we use DMRG to study superconducting tendencies driven by moderate and strong repulsive interactions. DMRG has been used with great success to explore a diverse selection of strongly correlated phenomena highlighted by stripes, spin-liquids, and superconductivity \cite{DMRGinAgeofMPS,KagomeHeisenberg,Stripes,StripesPairing,BergPDW,Dplusid,SpinLiquid,ChiralSpinLiquid,DwaveMetal,EntanglementEntropyDMRG,J1J2}. However, since DMRG is quasi-1D in nature, no true long-range order can be seen in the correlations. Thus in order to access our system's superconducting tendencies we implement a pair-edge-field motivated by the field-pinning approach first introduced in \cite{stripedttj} and employed similarly in \cite{VenderleyPDW}.
By studying how the system responds to the pair-edge field comparing responses in different superconducting channels, one can assess 
 the model's propensity for superconducting instabilities in different channels.

\section{Model and Method}
\begin{figure}[H]
\centering
\includegraphics[width=0.8\linewidth]{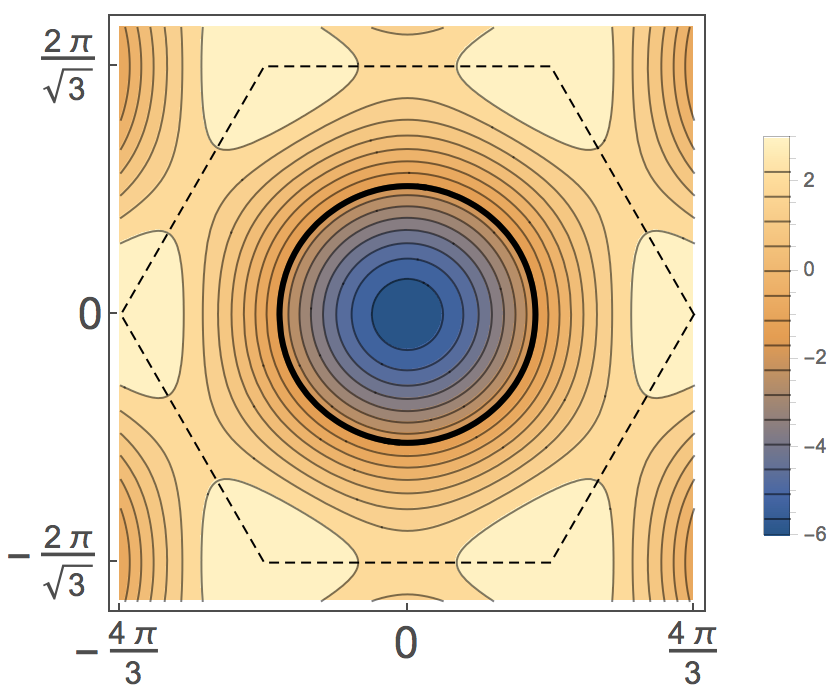}
\caption{Contour plot of the band structure. The Brillouin zone is marked by the dashed line. A typical Fermi surface for our hole-doped calculations is marked by the thick, black line, here $n=0.65$. The color legend indicates the energy of the band at a given k-point (for t=1).}
\label{fig:band_structure}
\end{figure}

The full model hamiltonian is
$H_{\rm tot}=H + H_{\rm edge}$, where $H$ is the standard Hubbard model on a triangular lattice:
\begin{equation}
H = -t \sum\limits_{\langle i, j \rangle \sigma} c^{\dagger}_{i \sigma} c_{j \sigma} - \mu \sum\limits_{i \sigma} n_{i \sigma} + U \sum\limits_{i} n_{i \uparrow} n_{i \downarrow},
\label{eq:H}
\end{equation}
 and $H_{\rm edge}$ is the pair-edge field. 
The non-interacting component of our base Hamiltonian $H$ has the spin-degenerate band structure shown in Fig. \ref{fig:band_structure}.  As evidenced by the figure, this model has a D$_6$ lattice symmetry and $T$-reversal symmetry. 
At half-filling, earlier studies found a metal - spin-ordered Mott insulator transition with increasing $U/t$ where the intermediate regime, $U/t \approx 8.3 - 10.6$ is a gapped spin-liquid \cite{Yunoki2017,Zaletel2018}.
Here we set the chemical potential $\mu$ such that the system is hole-doped away from half-filling. 

\begin{figure}[H]
\centering
\includegraphics[width=0.9\linewidth]{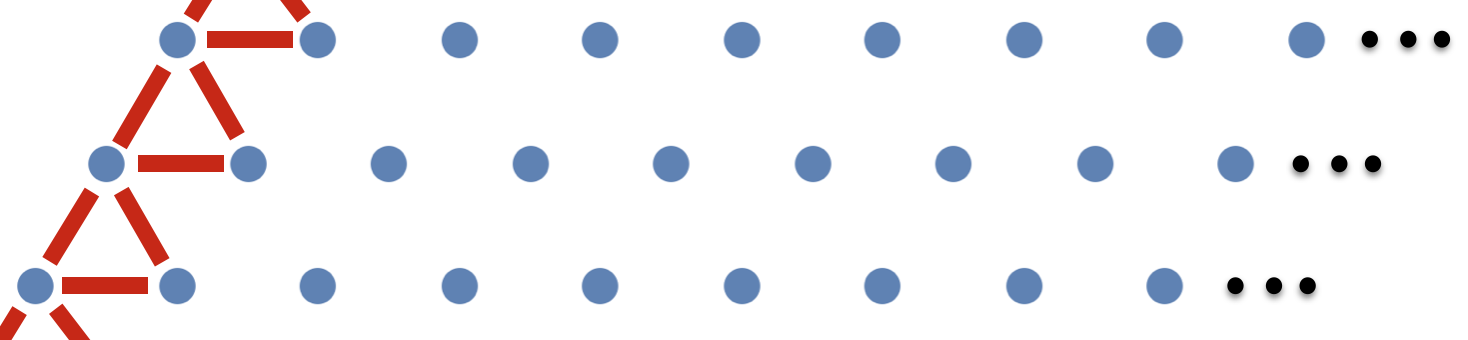}
\caption{A portion of our lattice with the location of the pair edge-fields marked in red.}
\label{fig:lattice}
\end{figure}

The edge field we apply, $H_{\rm edge}$, takes the form of 
\begin{equation}
  H_{edge} = \sum\limits_{\langle i, j \rangle \in \text{Edge}, \sigma \neq \sigma' }  V^{\sigma \sigma'}_{i j} c_{i \sigma} c_{j \sigma'} + h.c.  
\end{equation}
This edge field is depicted in Fig.~\ref{fig:lattice} as red lines connecting sites adjacent to the left edge of our lattice. As is standard for DMRG, our calculation is performed on a cylinder where we use L=18,24 along the open direction and W=3 for the periodic direction. The introduction of a pair-field breaks the $U(1)$ gauge symmetry and the associated particle number conservation as well as the D$_6$ lattice symmetry of the Hubbard model in Eq.~\eqref{eq:H}. Nevertheless, we can determine the susceptibility of the system to different pairing symmetries by observing the pairing response of the system under changes of 
the angular and spin dependence of $V^{\sigma \sigma'}_{i j}$.  
The notable irreducible representations of D$_6$ include two 2-dimensional irreps E$_1$ and E$_2$ corresponding to linear combinations of triplet p-wave (\textit{p}$\pm i$\textit{p}) and singlet d-wave  (\textit{d}$\pm i$\textit{d}) basis functions respectively. We will look into the amplitude and phase of singlet and triplet pair fields. 

For our DMRG simulations we utilize the ITensor library developed by Miles Stoudenmire and Steve White \cite{itensor}. We perform 14 sweeps with the final sweep containing up to 2500 states. The system is initialized by randomly sampling the even particle states, picking 10 states from each even particle sector and constructing a superposition of these states with coefficients randomly sampled from the interval [-1,1]. Proper implementation is ensured by double checking with exact diagonalization on small systems. 

Note that since our calculation does not conserve particle number, the filling in the converged system has a non-trivial dependence on the interaction strength. Consequently, working at a specified filling is difficult as it is not directly set by the chemical potential. Thus for simplicity we set $\mu = 0$ for all simulations which corresponds to fillings $n \sim 0.65$ for $U/t = 2$ and $n \sim 0.4$ for $U/t = 10$ where $n$ here is defined so that $n \in [0, 2]$ and half-filling corresponds to $n=1$. 

\section{Results}

We assess the superconducting tendency in a particular channel by measuring the bond pair order parameter $\Delta_{ij} = \langle c^{\dagger}_{i\uparrow}c^{\dagger}_{j\downarrow} \rangle$, defined below for the singlet and triplet channels: 
\begin{align}
\begin{split}
\Delta_{ij}^{singlet} &= \langle c^{\dagger}_{i\uparrow}c^{\dagger}_{j\downarrow} - c^{\dagger}_{i\downarrow}c^{\dagger}_{j\uparrow} \rangle \\
\Delta_{ij}^{triplet} &= \langle c^{\dagger}_{i\uparrow}c^{\dagger}_{j\downarrow} + c^{\dagger}_{i\downarrow}c^{\dagger}_{j\uparrow} \rangle.
\label{eq:orderparam}
\end{split}
\end{align}
In order to let inherent pairing symmetry emerge, we 
study the phase structure of the bond pair order parameters in response to a {\em random phase} singlet and triplet edge-field with constant amplitudes. This allows the dominant pairing symmetry (assuming one exists) to develop naturally from the RG flow of DMRG. We look for phase coherence over the extent of the system and discern the symmetry of the phase coherent channel. 
We then assess the strength of pairing tendency in each channel through the amplitude of the bond pair order parameter in response to the pair fields of $p$-wave and $d$-wave symmetries. We compare these responses in a system with moderately repulsive interaction of  $U/t = 2$ and $n \sim 0.65$ to those in a system with strong repulsive interaction of $U/t = 10$ and $n \sim 0.4$.

\begin{figure}[ht]
\centering
\subfigure[]{
    \raisebox{+0.1\height}{\includegraphics[width=0.9\linewidth]{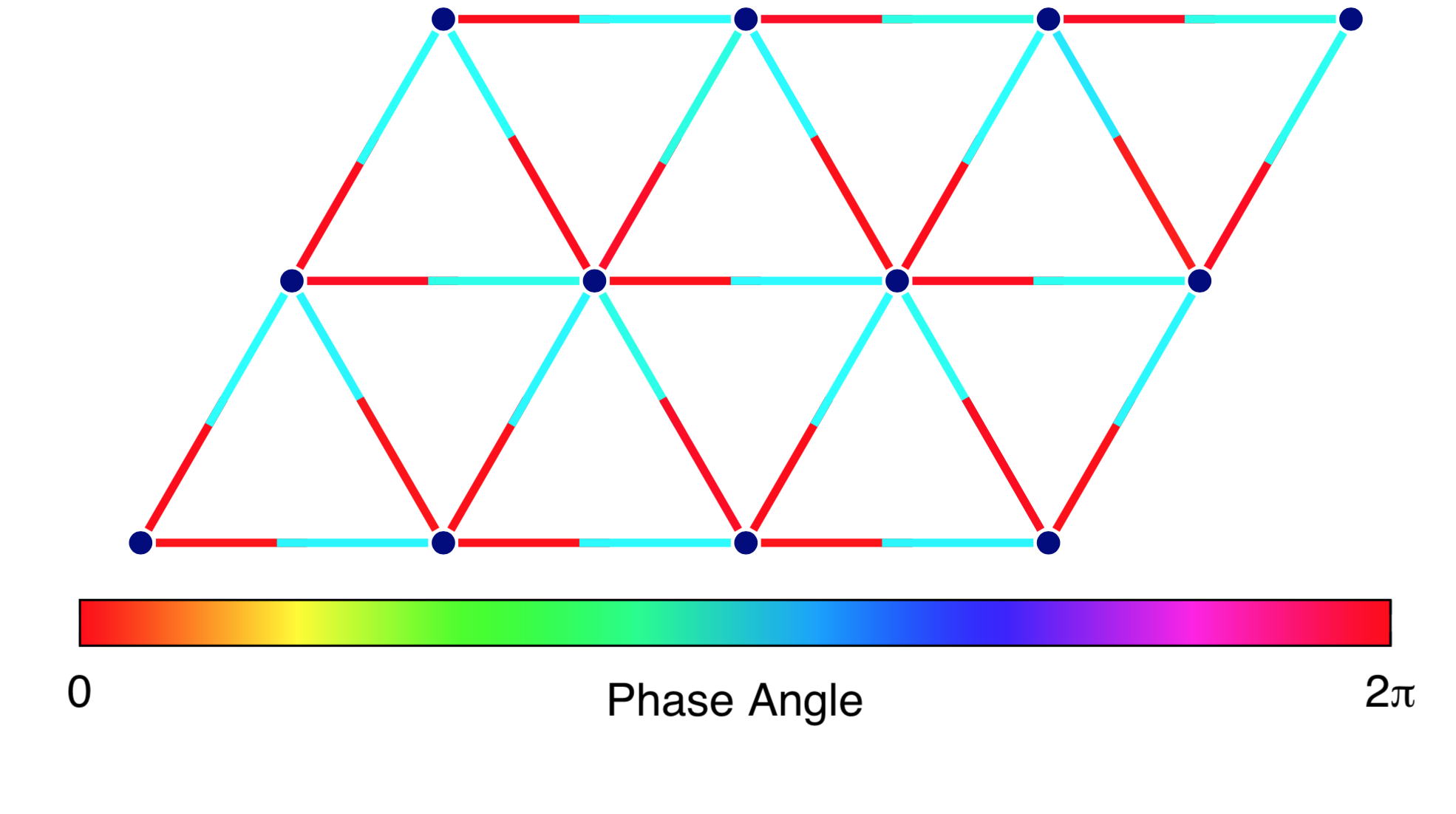}}
    \label{fig:phase_trip_rand_U2_trunc}}\\
\subfigure[]{
    \includegraphics[width=0.9\linewidth]{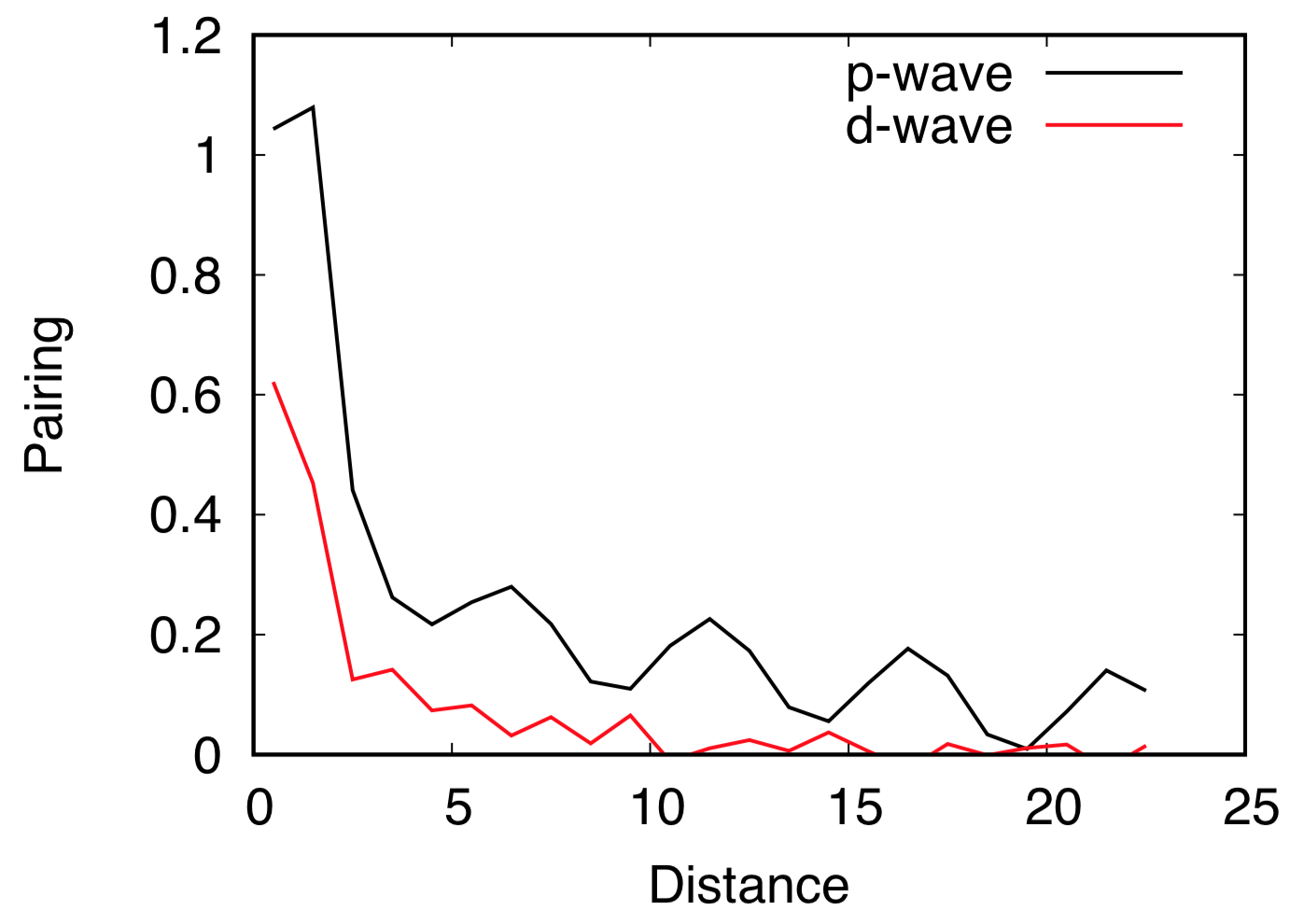}
    \label{fig:amp_trip_rand_U2}}
\caption{(a) SC phase plots for random edge-field, triplet channel, $U/t=2$ and $n \sim 0.65$. The phase for each $\Delta^{triplet}_{ij}$ is represented by the color of the bond $ij$. (b) Amplitude of the SC response with p-wave edge-field (black) and d-wave edge-field (red) for $U/t=2$. Only the parity channel corresponding to the parity of the edge field is shown. Reported amplitudes are normalized by the applied field ($V = 0.1$).}
\label{fig:SCamp}
\end{figure}

For moderate repulsion of $U/t = 2$ and $n \sim 0.65$, we find a strong superconducting response with $p$-wave symmetry in the triplet channel and much weaker and phase incoherent superconducting response in the singlet channel. 
The dominance of $p$-wave pairing response for this moderate repulsion is evident from the phase structure established for the much of the system in response to the random edge field. We show the phase of the triplet channel pair order parameter for $U/t=2$ from the segment of the system in the bulk in Fig.~\ref{fig:phase_trip_rand_U2_trunc}. 

Note that in this phase plot, the phase undergoes a $\pi$-phase shift in between the drawn bonds since $\Delta_{ij}^{triplet}$ is odd parity i.e. $\Delta_{ij}^{triplet} = -\Delta_{ji}^{triplet}$. 
Here, the induced pairing symmetry about any given lattice point is clearly p-wave. We suspect that nodal SC occurs as a result of finite size effects and that its chiral counterpart \textit{p}$\pm i$\textit{p} should emerge in the two-dimensional limit where the condensation energy is maximized by fully gapping the Fermi surface. For the phase map of entire system, see Fig.~\ref{fig:SC_full_phases} in the appendix. We also show phase disordered response in the singlet channel for $U/t=2$ in the appendix. 

\begin{figure}[H]
\centering
\subfigure[]{
    \raisebox{+0.1\height}{\includegraphics[width=0.9\linewidth]{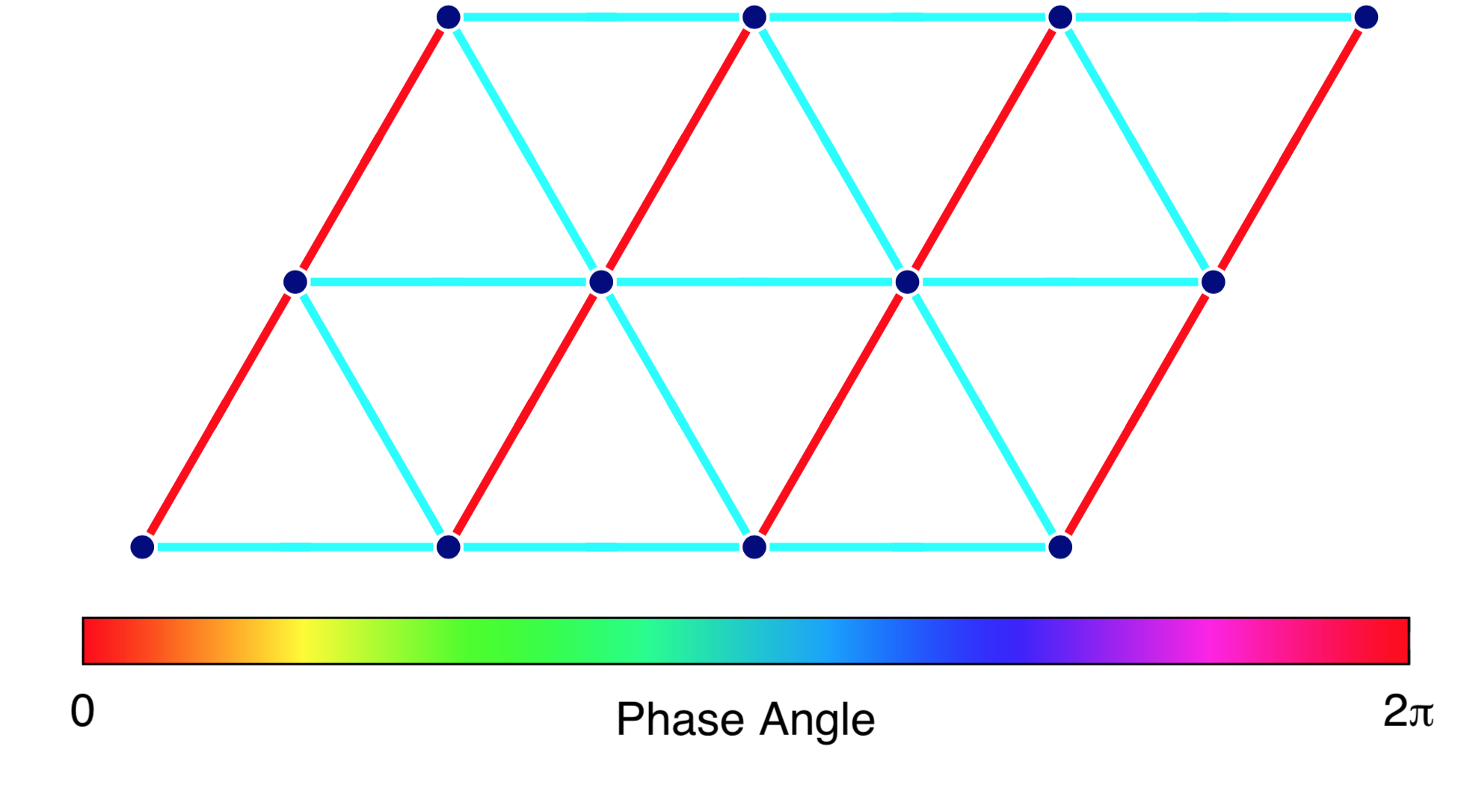}}
    \label{fig:phase_sing_rand_U10_trunc}}\\
\subfigure[]{
    \includegraphics[width=0.9\linewidth]{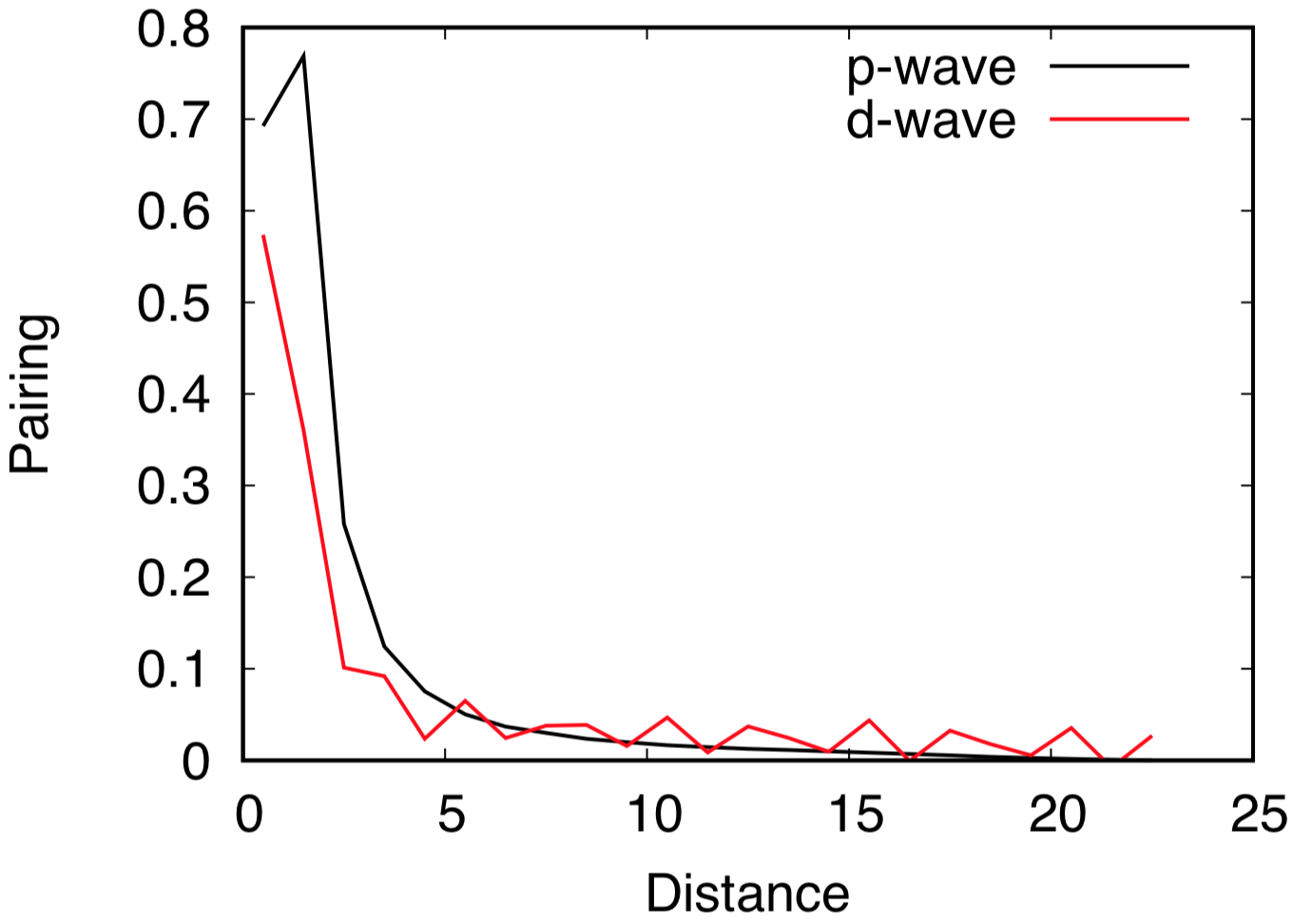}
    \label{fig:amp_sing_rand_U10}}
\caption{(a) SC phase plots for random edge-field, singlet channel, $U/t=10$ and $n \sim 0.4$. The phase for each $\Delta^{singlet}_{ij}$ is represented by the color of the bond connecting site i and j. (b) Amplitude of the SC response with p-wave edge-field (black) and d-wave edge-field (red) for $U/t=10$. Only the parity channel corresponding to the parity of the edge field is shown. Reported amplitudes are normalized by the applied field ($V = 0.1$).}
\label{fig:SCamp}
\end{figure}

Superconducting responses change qualitatively for strong repulsive interaction of $U/t = 10$ and $n \sim 0.4$. 
 Here we see a response that is symmetry distinct from the above moderate $U/t$ regime in response to the random phase edge fields. Namely,
the induced superconductivity in the triplet channel is weak and phase incoherent while that in the singlet channel is dominant with \textit{d}-wave symmetry. The truncated SC phase plot is given in Fig. \ref{fig:phase_sing_rand_U10_trunc}. Again, we expect \textit{d}$\pm i$\textit{d} in the 2D limit. The full phase map for both the singlet and triplet channel under random phase edge fields are given in  the appendix~\ref{fig:SC_full_phases}. 

In order to determine the strength of the response, we now turn away from the random edge-fields and instead employ pair-fields with \textit{p}-wave and \textit{d}-wave symmetries. Consistent with previous results, we find that for $U/t = 2$ and $n \sim 0.65$ the response in the \textit{p}-wave channel is much stronger that the \textit{d}-wave channel, see Fig. \ref{fig:amp_trip_rand_U2} while the opposite is true for $U/t = 10$ and $n \sim 0.4$ as seen in Fig. \ref{fig:amp_sing_rand_U10}.

\section{Conclusion}
Several recent works on this model at half-filling have focused on the realization of a chiral spin-liquid state at between the low $U/t$ metal and the high $U/t$ 120\degree N\'{e}el state\cite{Mishmash2015,Yunoki2017,Zaletel2018}. In light of these spin-liquid works, investigating superconductivity in this model under hole-doping is especially interesting since the hole-doping of a spin-liquid, particularly a chiral one, is thought to yield exotic superconductivity \cite{Anderson1987, Laughlin1987, Wen1989}. Our results are consistent with this picture, though the study of the doping dependence is warrented.
\footnote{Note that a previous study on this model \cite{Moreno2013} explored the regime $n \in [0.66, 1.0 ]$ where they found $\textit{d}\pm i\textit{d}$ for $U/t = 8.5$ and $n \in [0.7, 1.0 ]$.} 

The results discussed in this paper are 
particularly interesting 
in the context of on-going experimental efforts in twisted TMD bilayers. Not only the hetero-TMD bilayers are well-described by the triangular lattice Hubbard model\cite{MacDonald2018} the key parameter $U/t$ is highly tunable in these systems through the twist angle. 
Our results suggest that the strength of the repulsion and filling in these systems would play an integral role in determining the pairing symmetry. Specifically, weak to moderate repulsion will likely facilitate $\textit{p} \pm i \textit{p}$ SC while stronger repulsion will tend towards $\textit{d} \pm i \textit{d}$.

ACKNOWLEDGEMENTS: We thank Allan MacDonald, Leon Balents, Liang Fu, Ashvin Vishwanath, Mike Zaletel, Donna Sheng, and Garnet Chen for helpful discussions. We acknowledge support from the National Science Foundation through the Platform for the Accelerated Realization, Analysis, and Discovery of Interface Materials (PARADIM) under Cooperative Agreement No. DMR-1539918.

\bibliographystyle{apsrev4-1}
\bibliography{Tri_sources}

\appendix
\setcounter{figure}{0} \renewcommand{\thefigure}{A.\arabic{figure}}

\begin{widetext}

\section{Appendix: Phase Structure for full lattice}

We provide the simulation results depicting the phase structure induced by the random phase edge field for the entire lattice as opposed to the truncated versions provided in the main text. We also provide the singlet and triplet components for each interaction regime studied.

\begin{figure}[H]
\centering
\subfigure[]{
    \includegraphics[width=0.8\linewidth]{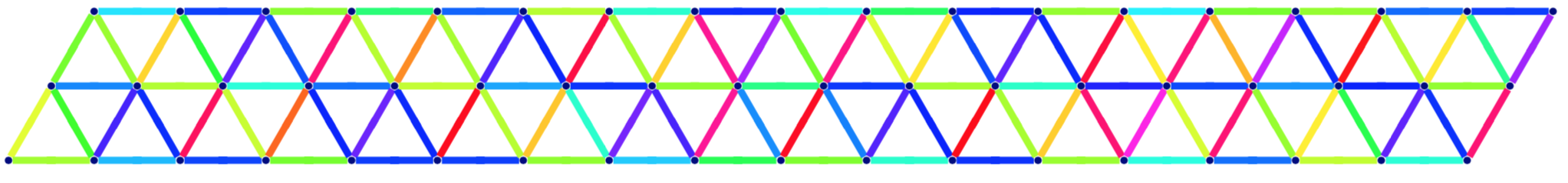}
    \label{fig:phase_sing_rand_U2}}
\subfigure[]{
    \includegraphics[width=0.8\linewidth]{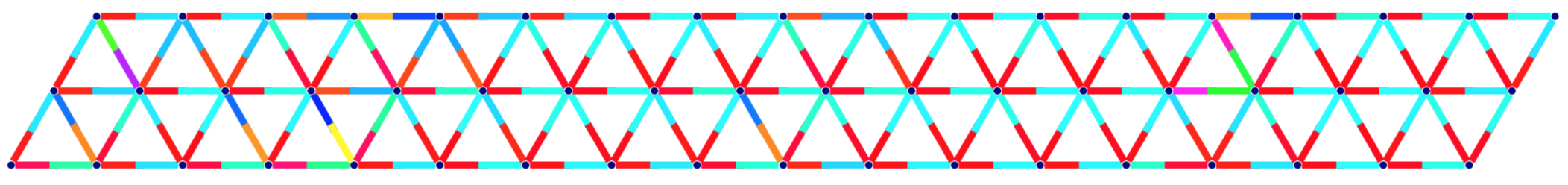}
    \label{fig:phase_trip_rand_U2}}
\subfigure[]{
    \includegraphics[width=0.8\linewidth]{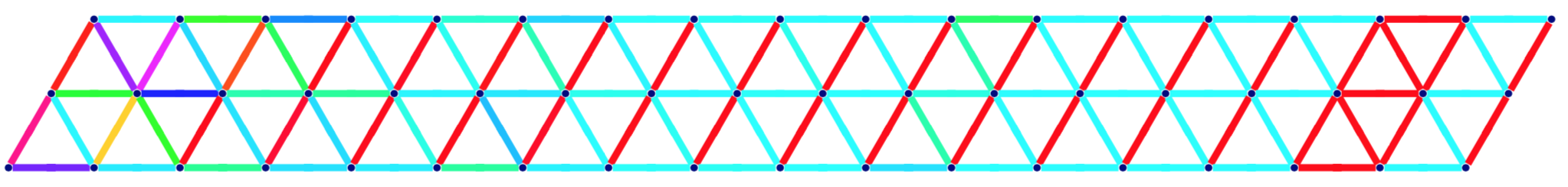}
    \label{fig:phase_sing_rand_U10}}
\subfigure[]{
    \includegraphics[width=0.8\linewidth]{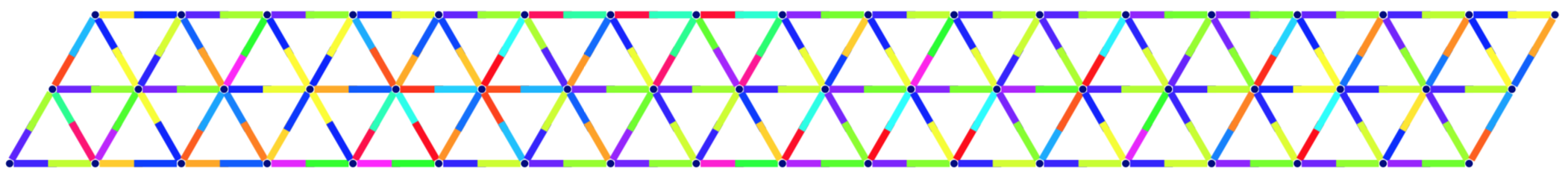}
    \label{fig:phase_trip_rand_U10}}
\subfigure{
    \includegraphics[width=0.8\linewidth]{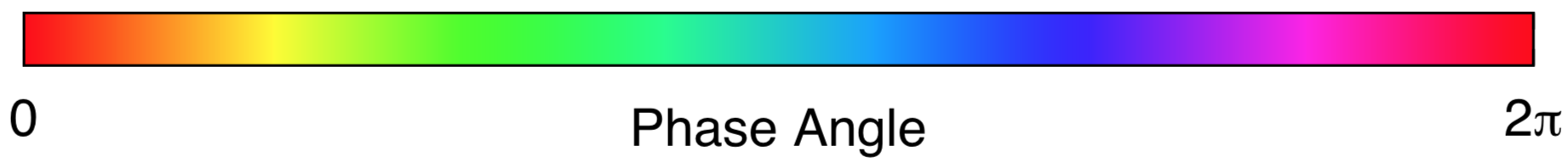}}
\caption{(a) SC phase plots for random edge-field, singlet channel, $U/t=2$ and $n \sim 0.65$ (b) SC phase plots for random edge-field, triplet channel, $U/t=2$ and $n \sim 0.65$ (c) SC phase plots for random edge-field, singlet channel, $U/t=10$ and $n \sim 0.4$ (d) SC phase plots for random edge-field, triplet channel, $U/t=10$ and $n \sim 0.4$.}
\label{fig:SC_full_phases}
\end{figure}

\end{widetext}

\end{document}